\documentstyle[preprint,revtex]{aps}
\tightenlines
\begin{document}
\draft
\begin{title}
Relativistic and Nuclear Structure Effects in \\
Parity-Violating Quasielastic Electron Scattering.
\end{title}
\author{C.J. Horowitz}
\begin{instit}
Department of Physics and Nuclear Theory Center, \\
Indiana University, Bloomington, Indiana 47405
\end{instit}
\author{J. Piekarewicz}
\begin{instit}
Supercomputer Computations Research Institute, \\
Florida State University, Tallahassee, FL 32306
\end{instit}
\begin{abstract}
The parity violating longitudinal asymmetry ${\cal A}$ is calculated
for quasielastic electron scattering.  We use a variety of relativistic
mean field models for the response of nuclear matter and $^{12}$C  at
a momentum transfer of q=550 MeV/c.  Relativistic effects from a reduced
nucleon mass, RPA correlations and vacuum polarization can all change
${\cal A}$ by a relatively large amount. These large nuclear structure
corrections could make it impossible to extract strange quark matrix
elements or radiative corrections to weak axial currents.
\end{abstract}
\newpage

\narrowtext

\section{Introduction}
\label{secintro}

Parity violating electron scattering probes weak neutral currents and
can provide very interesting information on strange quark matrix elements
in nucleons and nuclei.  One is interested in the form factors of the
the vector (both electric and magnetic) and the axial strange quark
currents in a nucleon. These three form factors contain new information
on nucleon structure which can help guide and constrain QCD based models.

It will take a number of measurements to determine separately all of
these form factors.  Furthermore, there are important complications
from radiative corrections~\cite{mushol90} which hinder the
extraction of strange quark matrix elements. Therefore, one
anticipates a program of several electron scattering experiments
involving proton and different nuclear targets. These results will
then be combined with neutrino scattering in order to extract the most
accurate strange-quark information.

Clearly one experiment is elastic electron proton scattering with
polarized electrons ($\vec e$,$p$). The parity violating analyzing
power (the cross section difference between positive and negative
helicity electrons over the sum) is sensitive to weak neutral currents.
Indeed, the SAMPLE experiment~\cite{sample89} will place (somewhat
crude) limits on the strange quark magnetic form factor.

Neutrino scattering may provide the best information on the axial-vector
form factor. For example, the BNL experiment measuring neutrino and
antineutrino proton scattering claimed to determine a nonzero strange
quark contribution~\cite{bnl91}. Note, for experimental reasons, most
neutrino experiments measure a combination of elastic scattering from
free protons and quasielastic scattering from nucleons bound in nuclei.
In a later paper we will examine nuclear structure corrections to
quasielastic neutrino scattering and how these could affect the extraction
of strange quark matrix elements.

Even if the strange axial current was accurately determined
(in neutrino scattering), uncertainties in axial currents could
still haunt electron scattering experiments. This is because of radiative
corrections to the axial current which are very different (larger) for
electrons than for neutrinos~\cite{mushol90}. For example, a small
``wrong'' parity admixture could exist in the nucleon (perhaps due to
weak couplings of the nucleon to its meson cloud).  The electron could
then scatter completely electromagnetically from the small wrong parity
admixture. These radiative corrections could be large (since they would
not involve the small weak vector coupling of the electron
$g_{\scriptscriptstyle V}^{\scriptscriptstyle e}$
(see Eq.~(\ref{atp}) and Table \ref{tablegs}).
Furthermore, some radiative corrections are difficult to calculate since
they involve unknown strong interaction matrix elements.

One alternative is to measure the radiative corrections by performing
another experiment (besides elastic electron-proton scattering)
which has a different combination of radiative corrections and strange
quark contributions. One appealing possibility is to measure parity
violation in quasielastic electron scattering.  This is favorable
experimentally because it has a large cross section and a relatively
large asymmetry. Indeed, there has already been one pioneering
measurement on ${}^{9}$Be~\cite{heil89}. [However, this experiment was
not accurate enough to determine strange quark matrix elements].

Because quasielastic scattering also involves neutrons one measures a
different isospin combination than in elastic proton scattering.  This
could allow one to separate isoscalar strange quark contributions from
radiative corrections to the isovector axial current.  With this in
mind, a parity violating quasielastic experiment has been proposed by
Donnelly and collaborators~\cite{donnel92,hadjim92}.

However if a nuclear target is involved, one must immediately address two
questions: (a) do unknown nuclear structure effects complicate the
extraction of strange quark matrix elements and (b) are the strange quark
matrix elements themselves modified in a nucleus?  We discuss the second
(very interesting) question first.

It is plausible that strange quark currents are different in a nucleus than
in a nucleon. For example, the old EMC experiment revealed that up and
down quark structure functions are changed by about 10\%. One might
expect the strange quark content to increase with density.  Indeed, in the
limit of very high density one expects a transition from nuclear to quark
matter.  Furthermore, it is believed that the ground state of quark matter
(in weak equilibrium) is strange matter containing a large number of
strange quarks. Here the large up and down quark Fermi energies overcome
the inhibiting effect of the strange quark mass. Even at much lower
densities, one can speculate that increasing the up and down quark Fermi
energies would help produce more virtual strange-quark pairs. Thus the
strange quark content could increase with density.

In a hadronic picture the same conclusion can be reached.  Calculations
for neutron matter suggest that the fraction of real hyperons increases
with density at high density~\cite{glemos91}. Furthermore, if a Kaon
condensate~\cite{taktos92} were to form at high density, the strange
quark content would be very large.
At lower densities, one could expect the interactions between nucleons and
the large nucleon Fermi energy will help drive virtual transitions of
nucleons to Kaons and hyperons. Thus the ``Kaon cloud'' in a nucleus could
be larger than in a nucleon. This would increase the strange quark content
of nuclear matter.

At present very little is known about the strange quark content of
nucleons much less that of a nucleus.  Therefore, we will make the minimal
assumption that the strange quark matrix elements are the same in a nucleus
as in a nucleon.  We leave the density dependence of the strange quark
contributions as a very interesting question for further work.

In this paper we address the first (more pedestrian but nevertheless
crucial) question.  How much do nuclear structure uncertainties complicate
the extraction of strange quark matrix elements from a quasielastic
electron scattering measurement?  Because the asymmetry at backward
angles involves a ratio of the parity violating weak response to a very
similar parity conserving electromagnetic response most nuclear structure
effects simply cancel in the ratio. We discuss this in detail below.
However, the weak vector current has a very different isospin character
from the electromagnetic current.  Therefore, one is sensitive to the
ratio of isoscalar to isovector responses.  A nuclear structure
effect that changes the ratio of isoscalar to isovector responses
will change the asymmetry and destroy ones ability to extract
strange quark matrix elements.

Donnely {\it et al.,} have argued that this isospin ratio is small and
under control (in the transverse response)~\cite{donnel92}.
[Note, the transverse response dominates over the longitudinal for
large electron scattering angles, see below]. Most isoscalar effects
are small because of the very small isoscalar anomalous moment of the
nucleon.  This argument is valid even if there are big random phase
approximations (RPA) or other nuclear structure corrections to the bare
isoscalar response. The overall isoscalar response is still expected to
be small because the electromagnetic probe only couples very weakly
through the small isoscalar anomalous moment.

However, there can be sizable contributions of the Dirac
${F}_{\scriptscriptstyle 1}$
form factor to the isoscalar response. In an earlier paper~\cite{horow92},
we argued that these contributions could be significantly enhanced by
relativistic effects. In the present paper we elaborate on these earlier
results and examine RPA and vacuum polarization corrections. Furthermore,
we extend earlier nuclear-matter results and present full finite-nucleus
responses.

Our formalism for parity violating electron scattering is presented in
Sec.~\ref{secformal} where the asymmetry is expressed as a ratio
of several nuclear response functions. Definitions of these response
functions are presented in Sec.~\ref{secnucresp} and results in
Sec.~\ref{secresults}. For simplicity we perform calculations
only for $^{12}$C and nuclear matter at a momentum transfer of
$q=550$~MeV/c. Sec.~\ref{secconcl} presents a summary of our
findings and our conclusions. We find that relativistic and other nuclear
structure effects are relatively large and that these may destroy the
interpretability of a parity violating quasielastic measurement.

\section{Formalism}
\label{secformal}

\subsection{Nuclear Response Functions}
\label{secnucresp}

In this section we express the parity violating asymmetry in
terms of a number of response functions for weak and electromagnetic
currents. This formalism has also been presented in
Refs.~\cite{donnel92,horow92,musolf92}.
The longitudinal analyzing power, or asymmetry, ${\cal A}$ is the
cross section difference for scattering electrons of positive and
negative helicity divided by the sum. It clearly vanishes if parity
is conserved. However, in the standard model, interference terms
between the weak neutral and the electromagnetic current produce a
small but nonzero ${\cal A}$.

In lowest order, the inclusive (polarized) cross section
 \begin{equation}
   {d^{\scriptscriptstyle 2}\sigma_{\scriptscriptstyle h}
    \over d\Omega^{'}dE^{'}} \simeq
    \Big[
     l_{\scriptscriptstyle \mu\nu}^{\scriptscriptstyle \gamma\gamma}
     W^{\scriptscriptstyle \mu\nu}_{\scriptscriptstyle \gamma\gamma} +
     a_{\scriptscriptstyle 0}\tau
     l_{\scriptscriptstyle \mu\nu}^{\scriptscriptstyle \gamma Z}
     W^{\scriptscriptstyle \mu\nu}_{\scriptscriptstyle \gamma Z}
    \Big] \;, \label{sigma}
 \end{equation}
contains the one-photon exchange contribution plus the
$\gamma-Z^{0}$ interference term (the purely weak component
is negligible and will not be considered further). The leptonic
contribution to the cross section is contained in the two
tensors
 \begin{mathletters}
  \begin{eqnarray}
   && l_{\scriptscriptstyle \mu\nu}^{\scriptscriptstyle \gamma\gamma}
\equiv
      l_{\scriptscriptstyle \mu\nu}                               =
      \left[
      k_{\scriptscriptstyle \mu}k^{'}_{\scriptscriptstyle \nu} +
      k^{'}_{\scriptscriptstyle \mu}k_{\scriptscriptstyle \nu} -
      g_{\scriptscriptstyle{\mu\nu}}(k \cdot k')    +
      ih\varepsilon_{\scriptscriptstyle \mu\nu\alpha\beta}
      k^{\scriptscriptstyle \alpha}k^{'{\scriptscriptstyle \beta}}
      \right] \;,                                     \\
   && l_{\scriptscriptstyle \mu\nu}^{\scriptscriptstyle \gamma Z}
\equiv
      g_{\scriptscriptstyle V}^{\scriptscriptstyle e} l_{ \mu\nu}     +
      g_{\scriptscriptstyle A}^{\scriptscriptstyle e} l_{ \mu;\nu 5}  =
     (g_{\scriptscriptstyle V}^{\scriptscriptstyle e} + h
      g_{\scriptscriptstyle A}^{\scriptscriptstyle e}) l_{ \mu\nu} \;.
   \label{eltens}
  \end{eqnarray}
 \end{mathletters}
Here $h$ is the electron helicity and
$g_{\scriptscriptstyle V}^{\scriptscriptstyle e}
(g_{\scriptscriptstyle A}^{\scriptscriptstyle e})$ is the
weak-vector (weak-axial) charge of the electron (see Table \ref{tablegs}).
The response of the nuclear target is contained in the two
hadronic tensors $W^{\scriptscriptstyle \mu\nu}_{\scriptscriptstyle
\gamma\gamma}$
and $W^{\scriptscriptstyle \mu\nu}_{\scriptscriptstyle \gamma Z}$. The
electromagnetic
tensor, defined by
  \begin{equation}
     W^{\scriptscriptstyle \mu\nu}_{\scriptscriptstyle \gamma\gamma}
      ({\bf q},\omega) =
      \sum_{n}
      \langle \Psi_{n} |
       \widehat{J}^{ \mu}_{\scriptscriptstyle EM}({\bf q})
      | \Psi_{0} \rangle
      \langle \Psi_{0} |
       \widehat{J}^{ \nu}_{\scriptscriptstyle EM}({\bf -q})
      | \Psi_{n} \rangle
      \delta (\omega-\omega_{\scriptscriptstyle n}) \;,
   \label{defemtens}
  \end{equation}
can be expressed, using current conservation, Lorentz covariance,
and parity invariance, in terms of two independent response functions
   \begin{equation}
     W^{\scriptscriptstyle \mu\nu}_{\scriptscriptstyle \gamma\gamma} =
     W_{ 1}
     \left[g^{\scriptscriptstyle{\mu\nu}}-
     {q^{\scriptscriptstyle \mu}q^{\scriptscriptstyle \nu} \over q^2} \right] +
     {W_{ 2} \over M_{\scriptscriptstyle T}^{ 2}} \
     \left[p^{\scriptscriptstyle \mu}-{(p \cdot q) \over q^2}
q^{\scriptscriptstyle \mu}\right]
     \left[p^{\scriptscriptstyle \nu}-{(p \cdot q) \over q^2}
q^{\scriptscriptstyle \nu}\right] \;,
    \label{emtens}
   \end{equation}
where $q^{\scriptscriptstyle \mu}\equiv(\omega,{\bf q})$ is the four-momentum
transfer to the target, and $p^{\scriptscriptstyle \mu}$ and
$M_{\scriptscriptstyle T}$ are
the four-momentum and rest mass of the nuclear target respectively.
Since the electromagnetic tensor is symmetric under the exchange of
Lorentz $(\mu \leftrightarrow \nu)$ indices, the electromagnetic
contribution to the cross section becomes independent of the electron
helicity. Furthermore, since weak-interaction effects on the unpolarized
cross section are negligible, the parity conserving contribution to the
cross section is entirely electromagnetic and is given by,
 \begin{equation}
   {1 \over \sigma_{\scriptscriptstyle M}}
   \left[{d^{\scriptscriptstyle 2}\sigma_{\scriptscriptstyle h}
    \over d\Omega^{'}dE^{'}}\right]_{\scriptscriptstyle {\rm pc}}=
    \left[
     v_{\scriptscriptstyle L}S_{\scriptscriptstyle L}(\omega,q) +
     v_{\scriptscriptstyle T}S_{\scriptscriptstyle T}(\omega,q)
    \right] \;, \label{sigmapc}
 \end{equation}
where $\sigma_{\scriptscriptstyle M}$ is the Mott cross section,
$v_{\scriptscriptstyle L}$ and $v_{\scriptscriptstyle T}$ are kinematical
factors,
  \begin{equation}
     v_{\scriptscriptstyle L} = {Q^4 \over {\bf q}^{4}}  \;;   \quad
     v_{\scriptscriptstyle T} = {Q^2 \over 2{\bf q}^{2}} +
                        \tan^{2}(\theta/2)    \;;   \quad
     Q^{2} \equiv {\bf q}^{2}-\omega^{2}      \;,
  \end{equation}
and $S_{\scriptscriptstyle L}$ and $S_{\scriptscriptstyle T}$ are the
longitudinal and transverse
nuclear response functions. These are related to $W_{ 1}$ and
$W_{ 2}$ by the following simple relations
 \begin{mathletters}
  \begin{eqnarray}
     S_{\scriptscriptstyle L} &\equiv& W^{\scriptscriptstyle
00}_{\scriptscriptstyle \gamma\gamma}
                   = {{\bf q}^{2} \over Q^2} \; W_{ 1}  +
                     {{\bf q}^{4} \over Q^4} \; W_{ 2}  \;, \\
     S_{\scriptscriptstyle T} &\equiv& W^{\scriptscriptstyle
11}_{\scriptscriptstyle \gamma\gamma} +
                            W^{\scriptscriptstyle
22}_{\scriptscriptstyle \gamma\gamma}
                   = -2W_{ 1}                           \;.
   \label{wtos}
  \end{eqnarray}
 \end{mathletters}

	The parity-violating electron asymmetry, driven entirely
by the electromagnetic-weak interference term, is contained in the
hadronic tensor $W^{ \mu\nu}_{ \gamma Z}$. In order
to express the electromagnetic-weak tensor in terms of Lorentz-invariant
response functions we write it in terms of vector-vector and vector-axial
contributions, {\it i.e.,}
   \begin{equation}
     W^{\scriptscriptstyle \mu\nu}_{\scriptscriptstyle \gamma Z} \equiv
     \widetilde{W}^{\scriptscriptstyle \mu\nu}_{\scriptscriptstyle \gamma Z} +
     \widetilde{W}^{\scriptscriptstyle \mu;\nu5}_{\scriptscriptstyle \gamma Z}
\;,
    \label{intens}
   \end{equation}
where $\widetilde{W}^{\scriptscriptstyle \mu\nu}_{\scriptscriptstyle \gamma Z}
      (\widetilde{W}^{\scriptscriptstyle \mu;\nu5}_{\scriptscriptstyle \gamma
Z})$
arises from the interference of the electromagnetic vector current with
the vector(axial) component of the weak current, {\it i.e.,}
  \begin{mathletters}
   \begin{eqnarray}
     \widetilde{W}^{ \mu\nu}_{ \gamma Z}
      ({\bf q},\omega) &=&
      \sum_{n}
      \langle \Psi_{n} |
       \widehat{J}^{ \mu}_{\scriptscriptstyle EM}({\bf q})
      | \Psi_{0} \rangle
      \langle \Psi_{0} |
       \widehat{J}^{ \nu}_{\scriptscriptstyle NC}({\bf -q})
      | \Psi_{n} \rangle
      \delta (\omega-\omega_{\scriptscriptstyle n}) \;,  \\
     \widetilde{W}^{ \mu;\nu5}_{ \gamma Z}
      ({\bf q},\omega) &=&
      \sum_{n}
      \langle \Psi_{n} |
       \widehat{J}^{ \mu}_{\scriptscriptstyle EM}({\bf q})
      | \Psi_{0} \rangle
      \langle \Psi_{0} |
       \widehat{J}^{ \nu 5}_{\scriptscriptstyle NC}({\bf -q})
      | \Psi_{n} \rangle
      \delta (\omega-\omega_{\scriptscriptstyle n}) \;.
    \label{defweaktens}
   \end{eqnarray}
  \end{mathletters}
In particular, the purely vector component has exactly the same
Lorentz structure as the electromagnetic tensor displayed in
Eq.(\ref{emtens}), {\it i.e.,}
   \begin{equation}
     \widetilde{W}^{\scriptscriptstyle \mu\nu}_{\scriptscriptstyle \gamma Z}=
     \widetilde{W}_{ 1}
     \left[g^{\scriptscriptstyle{\mu\nu}}-
     {q^{\scriptscriptstyle \mu}q^{\scriptscriptstyle \nu} \over q^2} \right] +
     {\widetilde{W}_{ 2} \over M_{\scriptscriptstyle T}^{ 2}} \
     \left[p^{\scriptscriptstyle \mu}-{(p \cdot q) \over q^2}
q^{\scriptscriptstyle \mu}\right]
     \left[p^{\scriptscriptstyle \nu}-{(p \cdot q) \over q^2}
q^{\scriptscriptstyle \nu}\right] \;.
    \label{wvtens}
   \end{equation}
The vector-axial tensor, on the other hand, must be written in terms
of the only available pseudotensor that one can construct from
$p^{\scriptscriptstyle \mu}$ and $q^{\scriptscriptstyle \mu}$, namely,
   \begin{equation}
      \widetilde{W}^{\scriptscriptstyle \mu;\nu5}_{\scriptscriptstyle \gamma
Z}=-i
     {\widetilde{W}_{ A} \over M_{\scriptscriptstyle T}^{ 2}}
      \varepsilon^{\scriptscriptstyle \mu\nu\alpha\beta}
       p_{\scriptscriptstyle \alpha}q_{\scriptscriptstyle \beta} \;.
    \label{watens}
   \end{equation}

The parity-violating part of the cross section can now be obtained
by performing the contraction of the leptonic tensor
$l_{\scriptscriptstyle \mu\nu}^{\scriptscriptstyle \gamma Z}$
(Eq.(\ref{eltens}))
with the hadronic tensor $W^{\scriptscriptstyle
\mu\nu}_{\scriptscriptstyle \gamma Z}$.
This yields,
 \begin{equation}
   {1 \over \sigma_{\scriptscriptstyle M}}
   \left[{d^{\scriptscriptstyle 2}\sigma_{\scriptscriptstyle h}
    \over d\Omega^{'}dE^{'}}\right]_{\scriptscriptstyle {\rm pv}}=
    a_{\scriptscriptstyle 0}\tau h
    \left[
     g_{\scriptscriptstyle A}^{\scriptscriptstyle e}
      \left(
       v_{\scriptscriptstyle L} \widetilde{S}_{\scriptscriptstyle L}(\omega,q)
+
       v_{\scriptscriptstyle T} \widetilde{S}_{\scriptscriptstyle
T}(\omega,q) \right) +
       g_{\scriptscriptstyle V}^{\scriptscriptstyle e}
       v_{\scriptscriptstyle T'}\widetilde{S}_{\scriptscriptstyle A}(\omega,q)
    \right] \;,
 \end{equation}
where
 \begin{equation}
     v_{\scriptscriptstyle T'}=
     \tan(\theta/2)
     \left[{Q^2 \over 2{\bf q}^{2}} + \tan^{2}(\theta/2)\right]^{1/2} \;,
 \end{equation}
and the three additional nuclear response functions are given by
 \begin{mathletters}
  \begin{eqnarray}
     \widetilde{S}_{\scriptscriptstyle L} &\equiv&
     \widetilde{W}^{\scriptscriptstyle 00}_{\scriptscriptstyle \gamma Z}=
       {{\bf q}^{2} \over Q^2}\;\widetilde{W}_{ 1} +
       {{\bf q}^{4} \over Q^4}\;\widetilde{W}_{ 2} \;, \\
     \widetilde{S}_{\scriptscriptstyle T} &\equiv&
     \widetilde{W}^{\scriptscriptstyle 11}_{\scriptscriptstyle \gamma Z}+
     \widetilde{W}^{\scriptscriptstyle 11}_{\scriptscriptstyle \gamma Z}=
       -2\;\widetilde{W}_{ 1}                      \;, \\
     \widetilde{S}_{\scriptscriptstyle A} &\equiv&
        i\left(
         \widetilde{W}^{\scriptscriptstyle 1;2(5)}_{\scriptscriptstyle \gamma
Z} -
         \widetilde{W}^{\scriptscriptstyle 2;1(5)}_{\scriptscriptstyle
\gamma Z}\right) =
        2{ |{\bf q}| \over M_{\scriptscriptstyle T}}\;\widetilde{W}_{ A} \;.
   \label{wtoss}
  \end{eqnarray}
 \end{mathletters}
The factor $a_{\scriptscriptstyle 0}\tau$ ($\tau\equiv Q^{2}/4M^{2}$)
sets the scale for the magnitude of the
parity-violating effects and is given by,
   \begin{equation}
     a_{\scriptscriptstyle 0} =
      -{G_{F}M^{2} \over \sqrt{2}\pi\alpha}
      \simeq -3.172\times10^{-4}   \;.
   \end{equation}
Notice that in the above expression only terms linear in the electron
helicity were kept. Terms independent of the electron helicity make a
negligible contribution to the parity-conserving cross section and were
neglected. The parity-violating asymmetry, defined as the difference of
helicity cross sections divided by their sum, is now given by
 \begin{equation}
     {\cal A} = {d\sigma_{\scriptscriptstyle \uparrow}    -
                 d\sigma_{\scriptscriptstyle \downarrow}  \over
                 d\sigma_{\scriptscriptstyle \uparrow}    +
                 d\sigma_{\scriptscriptstyle \downarrow}} =
                {\cal A}_{\scriptscriptstyle L}+
                {\cal A}_{\scriptscriptstyle T}+
                {\cal A}_{\scriptscriptstyle T'} \;,
  \label{assym}
 \end{equation}
where
 \begin{mathletters}
  \begin{eqnarray}
       {\cal A}_{\scriptscriptstyle L}/a_{\scriptscriptstyle 0} \tau  &=&
        g_{\scriptscriptstyle A}^{\scriptscriptstyle e} \;
       {v_{\scriptscriptstyle L} \widetilde{S}_{\scriptscriptstyle L} \over
        v_{\scriptscriptstyle L} S_{\scriptscriptstyle L} +
        v_{\scriptscriptstyle T} S_{\scriptscriptstyle T}} \;,
\\
      {\cal A}_{\scriptscriptstyle T}/a_{\scriptscriptstyle 0} \tau   &=&
        g_{\scriptscriptstyle A}^{\scriptscriptstyle e} \;
       {v_{\scriptscriptstyle T} \widetilde{S}_{\scriptscriptstyle T} \over
        v_{\scriptscriptstyle L} S_{\scriptscriptstyle L} +
        v_{\scriptscriptstyle T} S_{\scriptscriptstyle T}} \;,
\\
      {\cal A}_{\scriptscriptstyle T'}/a_{\scriptscriptstyle 0} \tau   &=&
        g_{\scriptscriptstyle V}^{\scriptscriptstyle e} \;
       {v_{\scriptscriptstyle T'} \widetilde{S}_{\scriptscriptstyle A} \over
        v_{\scriptscriptstyle L} S_{\scriptscriptstyle L} +
        v_{\scriptscriptstyle T} S_{\scriptscriptstyle T}} \;.
   \label{atp}
  \end{eqnarray}
 \end{mathletters}

The hadronic tensors defined above are intimately related to a fundamental
many-body operator, namely, the current-current correlation function or
polarization tensor. This relation can be illustrated, for example, in
the case of the timelike-timelike component of the electromagnetic tensor,
  \begin{equation}
      W^{\scriptscriptstyle 00}_{\scriptscriptstyle \gamma\gamma} =
      -{1 \over \pi}{\cal I}m
      \sum_{n}
      {\langle \Psi_{0} |
       \widehat{J}^{\scriptscriptstyle 0}_{\scriptscriptstyle EM}(-q)
      | \Psi_{n} \rangle
      \langle \Psi_{n} |
       \widehat{J}^{\scriptscriptstyle 0}_{\scriptscriptstyle EM}(q)
      | \Psi_{0} \rangle
      \over \omega-\omega_{\scriptscriptstyle n}+i\eta}
      =
      -{1 \over \pi}{\cal I}m \;
       \Pi^{\scriptscriptstyle 00}_{\scriptscriptstyle
\gamma\gamma}({\bf q},\omega) ,
  \end{equation}
where the polarization tensor,
$\Pi^{\scriptscriptstyle 00}_{\scriptscriptstyle \gamma\gamma}$, has
been introduced in the
last line. Consequently, all five nuclear response functions can be
written in terms of appropriate components of several polarization
tensors, {\it i.e.,}
 \begin{mathletters}
  \begin{eqnarray}
     S_{\scriptscriptstyle L}({\bf q},\omega) &=&
      -{1 \over \pi}{\cal I}m \;
       \Pi^{\scriptscriptstyle 00}_{\scriptscriptstyle
\gamma\gamma}({\bf q},\omega) \;,      \\
     S_{\scriptscriptstyle T}({\bf q},\omega) &=&
      -{1 \over \pi}
       {\cal I}m
       \left(
       \Pi^{\scriptscriptstyle 11}_{\scriptscriptstyle
\gamma\gamma}({\bf q},\omega) +
       \Pi^{\scriptscriptstyle 22}_{\scriptscriptstyle
\gamma\gamma}({\bf q},\omega)
       \right) \;, \\
     \widetilde{S}_{\scriptscriptstyle L}({\bf q},\omega) &=&
      -{1 \over \pi}{\cal I}m \;
       \Pi^{\scriptscriptstyle 00}_{\scriptscriptstyle \gamma Z}({\bf
q},\omega) \;,      \\
     \widetilde{S}_{\scriptscriptstyle T}({\bf q},\omega) &=&
      -{1 \over \pi}
       {\cal I}m
       \left(
       \Pi^{\scriptscriptstyle 11}_{\scriptscriptstyle \gamma Z}({\bf
q},\omega) +
       \Pi^{\scriptscriptstyle 22}_{\scriptscriptstyle \gamma Z}({\bf
q},\omega)
       \right) \;, \\
     \widetilde{S}_{\scriptscriptstyle A}({\bf q},\omega) &=&
      -{1 \over \pi}
       {\cal I}m
       \left(
       i\Pi^{\scriptscriptstyle 1;2(5)}_{\scriptscriptstyle \gamma
Z}({\bf q},\omega) -
       i\Pi^{\scriptscriptstyle 2;1(5)}_{\scriptscriptstyle \gamma
Z}({\bf q},\omega)
       \right) \;.
  \end{eqnarray}
 \end{mathletters}
The advantage of relating all nuclear response functions to a
many-body operator, {\it i.e.,} the polarization tensor,
is that the responses can be systematically computed using well-known
many-body techniques ({\it e.g.,} Feynman diagrams). In particular,
this will enable us to calculate a nuclear response that goes beyond the
simple impulse, or uncorrelated, response by including many-body correlation
effects.

The various nuclear response functions can, then, be calculated from
imaginary parts of polarization insertions given a model for: (a) the
hadronic weak and electromagnetic currents and (b) the nuclear structure.
Hadronic currents are discussed in the next section. Subsequently,
we calculate the nuclear responses in a number of relativistic
mean-field models with and without RPA correlations and with various
treatments of vacuum polarization.

\subsection{Electromagnetic and Weak Hadronic Currents}
\label{seccurr}

	In this section we describe the electromagnetic and
weak hadronic currents that will be used in calculating the
five nuclear response functions. We start by writing the
electromagnetic current for the three lightest $(u, d, s)$ quarks
 \begin{equation}
    J^{ \mu}_{\scriptscriptstyle EM}=
      e_{\scriptscriptstyle u} \bar{u}\gamma^{ \mu}u +
      e_{\scriptscriptstyle d} \bar{d}\gamma^{ \mu}d +
      e_{\scriptscriptstyle s} \bar{s}\gamma^{ \mu}s \;.
 \end{equation}
For two quark flavors ($u$ and $d$) one customarily rewrites
the electromagnetic current in terms of isoscalar and isovector
contributions. The analogous expression in the three-flavor case
is given in terms of two octet (3 and 8) and the singlet SU(3)
currents, {\it i.e.,}
  \begin{eqnarray}
    J^{ \mu}_{\scriptscriptstyle EM} &=&
      e_{\scriptscriptstyle 0} \bar{q}\gamma^{ \mu}q    +
      e_{\scriptscriptstyle 3} \bar{q}\gamma^{ \mu}
                    {\lambda_{\scriptscriptstyle 3} \over 2}q +
      e_{\scriptscriptstyle 8} \bar{q}\gamma^{ \mu}
                    {\lambda_{\scriptscriptstyle 8} \over 2}q   \nonumber \\
                &=& \bar{q}\gamma^{ \mu}
                    {\lambda_{\scriptscriptstyle 3} \over 2}q +
    {1\over\sqrt{3}}\;\bar{q}\gamma^{ \mu}
                    {\lambda_{\scriptscriptstyle 8} \over 2}q   \;.
   \label{jem}
  \end{eqnarray}
where we have defined (and used) singlet and octet electromagnetic charges
 \begin{mathletters}
  \begin{eqnarray}
      e_{\scriptscriptstyle 0} &=&
     (e_{\scriptscriptstyle u}+e_{\scriptscriptstyle
d}+e_{\scriptscriptstyle s})/3          \;, \\
      e_{\scriptscriptstyle 3} &=&
     (e_{\scriptscriptstyle u}-e_{\scriptscriptstyle d})
          \;, \\
      e_{\scriptscriptstyle 8} &=&
     (e_{\scriptscriptstyle u}+e_{\scriptscriptstyle
d}-2e_{\scriptscriptstyle s})/\sqrt{3}  \;.
   \label{emcharges}
  \end{eqnarray}
 \end{mathletters}

	Single-nucleon electromagnetic form factors are now obtained
by evaluating the electromagnetic current between nucleon states {\it i.e.,}
 \begin{mathletters}
  \begin{eqnarray}
    \langle N(p's't')|
      J^{ \mu}_{\scriptscriptstyle EM} (T&=&0)
    | N(pst) \rangle = {1 \over \sqrt{3}}
    \langle N(p's't')|
      \bar{q}\gamma^{ \mu}
      {\lambda_{\scriptscriptstyle 8} \over 2}q
    | N(pst) \rangle \nonumber \\ &=&
    \bar{U}(p's') \left[
      F_{\scriptscriptstyle 1}^{\scriptscriptstyle (0)}
\gamma^{\scriptscriptstyle \mu} +
     iF_{\scriptscriptstyle 2}^{\scriptscriptstyle (0)}
\sigma^{\scriptscriptstyle \mu\nu}
                                  {q_{\scriptscriptstyle \nu}\over 2M}
    \right] U(p,s) \delta_{\scriptscriptstyle t't} \;,
    \label{emformfa}                    \\
    \langle N(p's't')|
      J^{ \mu}_{\scriptscriptstyle EM} (T&=&1)
    | N(pst) \rangle =
    \langle N(p's't')|
      \bar{q}\gamma^{ \mu}
      {\lambda_{\scriptscriptstyle 3} \over 2}q
    | N(pst) \rangle \nonumber \\ &=&
    \bar{U}(p's') \left[
      F_{\scriptscriptstyle 1}^{\scriptscriptstyle (1)}
\gamma^{\scriptscriptstyle \mu} +
     iF_{\scriptscriptstyle 2}^{\scriptscriptstyle (1)}
\sigma^{\scriptscriptstyle \mu\nu}
                                  {q_{\scriptscriptstyle \nu}\over 2M}
    \right] U(p,s) (\tau_{\scriptscriptstyle 3})_{\scriptscriptstyle t't} \;.
   \label{emformfb}
  \end{eqnarray}
 \end{mathletters}

	The weak-vector and weak-axial neutral currents can be written
in analogy to Eq.~(\ref{jem}). First, however, one rewrites the
SU(3)-singlet current in terms of one  octet (eight) and a pure
strange-quark current. This yields,
 \begin{equation}
    J^{ \mu}_{\scriptscriptstyle NC}=
      \xi_{\scriptscriptstyle V}^{\scriptscriptstyle (1)}
          \bar{q}\gamma^{ \mu}
          {\lambda_{\scriptscriptstyle 3} \over 2}q +
      \xi_{\scriptscriptstyle V}^{\scriptscriptstyle (0)}
          \bar{q}\gamma^{ \mu}
          {\lambda_{\scriptscriptstyle 8} \over 2}q +
      \xi_{\scriptscriptstyle V}^{\scriptscriptstyle (s)}
          \bar{s}\gamma^{ \mu}s  \;,
  \label{jvnc}
 \end{equation}
where the couplings are simple linear combinations of quark
weak-vector charges (see Table \ref{tablexis}), {\it i.e.,}
  \begin{mathletters}
   \begin{eqnarray}
      \xi_{\scriptscriptstyle V}^{\scriptscriptstyle (0)} &=& \sqrt{3} \;
       \left(
        g_{\scriptscriptstyle V}^{\scriptscriptstyle
u}+g_{\scriptscriptstyle V}^{\scriptscriptstyle d}
       \right) \;, \\
      \xi_{\scriptscriptstyle V}^{\scriptscriptstyle (1)} &=&
       \left(
        g_{\scriptscriptstyle V}^{\scriptscriptstyle
u}-g_{\scriptscriptstyle V}^{\scriptscriptstyle d}
       \right) \;, \\
      \xi_{\scriptscriptstyle V}^{\scriptscriptstyle (s)} &=&
       \left(
        g_{\scriptscriptstyle V}^{\scriptscriptstyle
u}+g_{\scriptscriptstyle V}^{\scriptscriptstyle d}+
        g_{\scriptscriptstyle V}^{\scriptscriptstyle s}
       \right) \;.
    \label{xicharge}
   \end{eqnarray}
  \end{mathletters}
Writing the hadronic current as above, enables one to express nucleon
matrix elements of the weak-vector current in terms of
electromagnetic isoscalar and isovector form factors plus a
strange-quark contribution (see Eqs.~(\ref{emformfa},~\ref{emformfa}))  ,
 \begin{mathletters}
  \begin{eqnarray}
    \langle N'|
      J^{ \mu}_{\scriptscriptstyle NC}(T=0)
    | N \rangle &=&
    \sqrt{3}\;\xi_{\scriptscriptstyle V}^{\scriptscriptstyle (0)}
    \langle N'|
      J^{ \mu}_{\scriptscriptstyle EM}(T=0)
    | N \rangle  +
    \xi_{\scriptscriptstyle V}^{\scriptscriptstyle (s)}
    \langle N'|
      \bar{s}\gamma^{\scriptscriptstyle \mu}s
    | N \rangle  \;,   \\
    \langle N'|
      J^{ \mu}_{\scriptscriptstyle NC}(T=1)
    | N \rangle &=&
    \xi_{\scriptscriptstyle V}^{\scriptscriptstyle (1)}
    \langle N'|
      J^{ \mu}_{\scriptscriptstyle EM}(T=1)
    | N \rangle  \;,
  \end{eqnarray}
 \end{mathletters}
where we have assumed that matrix elements of the strange-quark
current are equal for protons and neutrons. Defining single-nucleon
strange $({F}_{\scriptscriptstyle 1}^{\scriptscriptstyle (s)},
          {F}_{\scriptscriptstyle 2}^{\scriptscriptstyle (s)})$ and
weak-vector form factors
$(\widetilde{F}_{\scriptscriptstyle 1}^{\scriptscriptstyle (0)},
  \widetilde{F}_{\scriptscriptstyle 2}^{\scriptscriptstyle (0)},
  \widetilde{F}_{\scriptscriptstyle 1}^{\scriptscriptstyle (1)},
  \widetilde{F}_{\scriptscriptstyle 2}^{\scriptscriptstyle (1)})$
in analogy to Eqs. (\ref{emformfa},~\ref{emformfb}) we arrive at the
following useful relations between single-nucleon form factors
 \begin{mathletters}
  \begin{eqnarray}
   \widetilde{F}_{\scriptscriptstyle 1}^{\scriptscriptstyle (0)} &=&
   \sqrt{3}\;\xi_{\scriptscriptstyle V}^{\scriptscriptstyle (0)}
             {F}_{\scriptscriptstyle 1}^{\scriptscriptstyle (0)} +
             \xi_{\scriptscriptstyle V}^{\scriptscriptstyle (s)}
             {F}_{\scriptscriptstyle 1}^{\scriptscriptstyle (s)} \;, \qquad
   \widetilde{F}_{\scriptscriptstyle 1}^{\scriptscriptstyle (1)}  =
             \xi_{\scriptscriptstyle V}^{\scriptscriptstyle (1)}
             {F}_{\scriptscriptstyle 1}^{\scriptscriptstyle (1)} \;,
   \label{ftildea}                         \\
   \widetilde{F}_{\scriptscriptstyle 2}^{\scriptscriptstyle (0)} &=&
   \sqrt{3}\;\xi_{\scriptscriptstyle V}^{\scriptscriptstyle (0)}
             {F}_{\scriptscriptstyle 2}^{\scriptscriptstyle (0)} +
             \xi_{\scriptscriptstyle V}^{\scriptscriptstyle (s)}
             {F}_{\scriptscriptstyle 2}^{\scriptscriptstyle (s)} \;, \qquad
   \widetilde{F}_{\scriptscriptstyle 2}^{\scriptscriptstyle (1)}  =
             \xi_{\scriptscriptstyle V}^{\scriptscriptstyle (1)}
             {F}_{\scriptscriptstyle 2}^{\scriptscriptstyle (1)} \;.
   \label{ftildeb}
  \end{eqnarray}
 \end{mathletters}

Finally, one defines, in analogy to Eq.~(\ref{jvnc}), the weak-axial
neutral current in terms of octet and strange axial-vector currents
and the appropriate weak-axial charges (see Table~\ref{tablexis})
 \begin{equation}
    J^{ \mu 5}_{\scriptscriptstyle NC}=
      \xi_{\scriptscriptstyle A}^{\scriptscriptstyle (1)}
          \bar{q}\gamma^{ \mu}\gamma^{ 5}
          {\lambda_{\scriptscriptstyle 3} \over 2}q +
      \xi_{\scriptscriptstyle A}^{\scriptscriptstyle (0)}
          \bar{q}\gamma^{ \mu}\gamma^{ 5}
          {\lambda_{\scriptscriptstyle 8} \over 2}q +
      \xi_{\scriptscriptstyle A}^{\scriptscriptstyle (s)}
          \bar{s}\gamma^{ \mu}\gamma^{ 5}s  \;.
  \label{janc}
 \end{equation}
Single-nucleon form factors are obtained (neglecting induced-pseudoscalar
contributions) by evaluating octet and strange axial-vector currents between
nucleon states
 \begin{mathletters}
  \begin{eqnarray}
    \langle N(p's't')|
      \bar{q}\gamma^{ \mu}\gamma^{ 5}
      {\lambda_{\scriptscriptstyle 3} \over 2}q
    | N(pst) \rangle &=&
      G_{\scriptscriptstyle A}^{\scriptscriptstyle (3)}
    \left[ \bar{U}(p's')
      \gamma^{\scriptscriptstyle \mu}\gamma^{\scriptscriptstyle 5}
     U(p,s)\right] (\tau_{\scriptscriptstyle 3})_{\scriptscriptstyle t't} \;,
\\
    \langle N(p's't')|
      \bar{q}\gamma^{ \mu}\gamma^{ 5}
      {\lambda_{\scriptscriptstyle 8} \over 2}q
    | N(pst) \rangle &=&
      G_{\scriptscriptstyle A}^{\scriptscriptstyle (8)}
    \left[\bar{U}(p's')
      \gamma^{\scriptscriptstyle \mu}\gamma^{\scriptscriptstyle 5}
     U(p,s)\right] \delta_{\scriptscriptstyle t't}             \;,  \\
    \langle N(p's't')|
      \bar{s}\gamma^{ \mu}\gamma^{ 5}s
    | N(pst) \rangle &=&
      G_{\scriptscriptstyle A}^{\scriptscriptstyle (s)}
    \left[\bar{U}(p's')
      \gamma^{\scriptscriptstyle \mu}\gamma^{\scriptscriptstyle 5}
     U(p,s)\right] \delta_{\scriptscriptstyle t't}             \;.
  \end{eqnarray}
 \end{mathletters}
Isoscalar and isovector single-nucleon axial form factors are now obtained
from the following relations
 \begin{equation}
  \widetilde{G}_{\scriptscriptstyle A}^{\scriptscriptstyle (0)} =
            \xi_{\scriptscriptstyle A}^{\scriptscriptstyle (0)}
            {G}_{\scriptscriptstyle A}^{\scriptscriptstyle (8)} +
            \xi_{\scriptscriptstyle A}^{\scriptscriptstyle (s)}
            {G}_{\scriptscriptstyle A}^{\scriptscriptstyle (s)} \;, \qquad
  \widetilde{G}_{\scriptscriptstyle A}^{\scriptscriptstyle (1)} =
            \xi_{\scriptscriptstyle A}^{\scriptscriptstyle (1)}
            {G}_{\scriptscriptstyle A}^{\scriptscriptstyle (3)} \;.
 \end{equation}

In conclusion, the electromagnetic and weak-neutral currents that one is
to employ in calculating all nuclear response functions have been written
in terms of appropriate single-nucleon form factors
(described in the appendix)
and are given by,
  \begin{mathletters}
   \begin{eqnarray}
      \widehat{J}^{ \mu}_{\scriptscriptstyle EM} &=&
      \left[
        F_{\scriptscriptstyle 1}^{\scriptscriptstyle (0)}
        \gamma^{\scriptscriptstyle \mu} +
       iF_{\scriptscriptstyle 2}^{\scriptscriptstyle (0)}
        \sigma^{\scriptscriptstyle \mu\nu}{q_{\scriptscriptstyle \nu}\over 2M}
      \right] +
      \left[
        F_{\scriptscriptstyle 1}^{\scriptscriptstyle (1)}
        \gamma^{\scriptscriptstyle \mu} +
       iF_{\scriptscriptstyle 2}^{\scriptscriptstyle (1)}
        \sigma^{\scriptscriptstyle \mu\nu}{q_{\scriptscriptstyle \nu}\over 2M}
      \right] \tau_{\scriptscriptstyle z}  \;,
      \label {jelec}              \\
     \widehat{J}^{ \mu}_{\scriptscriptstyle NC} &=&
      \left[
        \widetilde{F}_{\scriptscriptstyle 1}^{\scriptscriptstyle (0)}
        \gamma^{\scriptscriptstyle \mu} +
       i\widetilde{F}_{\scriptscriptstyle 2}^{\scriptscriptstyle (0)}
        \sigma^{\scriptscriptstyle \mu\nu}{q_{\scriptscriptstyle \nu}\over 2M}
      \right] +
      \left[
        \widetilde{F}_{\scriptscriptstyle 1}^{\scriptscriptstyle (1)}
        \gamma^{\scriptscriptstyle \mu} +
       i\widetilde{F}_{\scriptscriptstyle 2}^{\scriptscriptstyle (1)}
        \sigma^{\scriptscriptstyle \mu\nu}{q_{\scriptscriptstyle \nu}\over 2M}
      \right] \tau_{\scriptscriptstyle z}  \;,
      \label {jweakv}           \\
     \widehat{J}^{ \mu5}_{\scriptscriptstyle NC} &=&
        \widetilde{G}_{\scriptscriptstyle A}^{\scriptscriptstyle (0)}
        \gamma^{\scriptscriptstyle \mu}\gamma^{\scriptscriptstyle 5} +
        \widetilde{G}_{\scriptscriptstyle A}^{\scriptscriptstyle (1)}
        \gamma^{\scriptscriptstyle \mu}\gamma^{\scriptscriptstyle 5}
        \tau_{\scriptscriptstyle z}  \;.
      \label{jweaka}
   \end{eqnarray}
  \end{mathletters}
The above equations are a major assumption of our model and prescribe how
the hadronic currents are taken ``off-shell''. We also assume that the form
factors are unchanged in the medium.

\subsection{Relativistic Mean-Field Responses}
\label{secmf}

In this section we describe the model used in the calculation
of the various weak and electromagnetic response functions.
Our starting point is a relativistic mean field approximation
to the Walecka model where nucleons move in strong scalar
$\Sigma_{\rm s}$ and vector $\Sigma_{\rm v}$ mean
fields~\cite{waleck74,serwal86}. In particular, the strong scalar
field is responsible for reducing the nucleon mass from its free-space
value to $M^*$
  \begin{equation}
    M^{*} \equiv M-\Sigma_{\rm s} \;.
  \end{equation}
For nuclear matter this is a constant that must be self-consistently
determined at every given density.  In a finite nucleus, however,
the effective mass depends on position. It is smaller at the center
and rises back to its free value $M$ at the surface.

These mean field models provide perhaps the minimal relativistic
description of a nuclear target.  The strong mean fields are very
closely related to the spin-orbit potential of the shell model.
Furthermore, $M^*$\ provides a minimal description of the binding
energy shift seen in the position of the quasielastic
peak~\cite{rosen80}.

We illustrate the formalism by presenting electromagnetic,
weak vector and weak axial response functions for nuclear matter.
Most of the formalism, however, remains valid in the finite
system. The appropriate modifications to be carried out for
a finite nucleus and for including RPA correlations will
be discussed at the end of the section.

The basic ingredient in calculating all nuclear response functions
is the nucleon Green's function (or Feynman propagator). The Green's
function describes the propagation of nucleons in the mean fields.
In a self-consistent Hartree approximation the nucleon propagator
is given by~\cite{serwal86}
  \begin{equation}
    G(k) = \left[ \rlap/{k} + M^* \right]
           \left(
             {1\over k^2-M^{*2}+i\epsilon}+
             i{\pi\over E^{*}_{k}}
             \delta(k^{0}-E^{*}_{k})
             \Theta(k_{\scriptscriptstyle F}-|{\bf k}|)
            \right) \equiv
            G_{\scriptscriptstyle {\rm F}}+G_{\scriptscriptstyle {\rm D}} \;,
   \label{greens}
  \end{equation}
where
$E^{*}_{k}=\sqrt{{\bf k}^{2}+M^{*2}}$\ and $k_{\scriptscriptstyle F}$ is the
Fermi momentum. $G_{\scriptscriptstyle {\rm F}}$ is that part of the propagator
having the same analytic structure as the free Feynman propagator.
The density-dependent part of the propagator, $G_{\scriptscriptstyle {\rm D}}$,
corrects $G_{\scriptscriptstyle {\rm F}}$ for the presence of occupied states
below the Fermi surface.

The electromagnetic (Eq.(\ref{jelec})) and weak-vector currents
(Eq.(\ref{jweakv})) have been written in terms of Dirac and Pauli
(or anomalous) contributions. This implies, that the electromagnetic
($\Pi^{\scriptscriptstyle \mu\nu}_{\scriptscriptstyle \gamma\gamma}$) as well
as the
weak-vector
($\Pi^{\scriptscriptstyle \mu\nu}_{\scriptscriptstyle \gamma Z}$) polarization
tensors will have to be evaluated in terms of individual polarizations
having Lorentz vector v=$\gamma^{\scriptscriptstyle \mu}$ and
tensor t$=\sigma^{\scriptscriptstyle \mu\nu}$ vertices, {\it i.e.,}
(ignoring isospin labels)
  \begin{mathletters}
   \begin{eqnarray}
     i\Pi^{\scriptscriptstyle \mu\nu}_{\scriptscriptstyle {\rm vv}} &=&
     \int {d^4k\over (2\pi)^4} {\rm Tr}
     \left[
       G(k)
       \gamma^{\scriptscriptstyle \mu}
       G(k+q)
       \gamma^{\scriptscriptstyle \nu}
      \right] \;,
      \label{pivv}     \\
     i\Pi^{\scriptscriptstyle \mu\nu}_{\scriptscriptstyle {\rm vt}} &=&
     \int {d^4k\over (2\pi)^4} {\rm Tr}
     \left[
       G(k)
       \gamma^{\scriptscriptstyle \mu}
       G(k+q)
       {i\sigma^{\scriptscriptstyle \nu\rho} q_{\scriptscriptstyle \rho}
        \over 2M}
      \right] \;,
      \label{pivt}     \\
     i\Pi^{\scriptscriptstyle \mu\nu}_{\scriptscriptstyle {\rm tt}} &=&
     \int {d^4k\over (2\pi)^4} {\rm Tr}
     \left[
       G(k)
       {\sigma^{\scriptscriptstyle \mu\eta} q_{\scriptscriptstyle \eta}\over
2M}
       G(k+q)
       {\sigma^{\scriptscriptstyle \nu\rho} q_{\scriptscriptstyle \rho}\over
2M}
      \right] \;.
      \label{pitt}
   \end{eqnarray}
  \end{mathletters}
Note that, both, the electromagnetic and the weak-vector
responses are driven by the same polarizations. The only
difference between the two responses arises from the use
of either electromagnetic or weak-vector single-nucleon form
factors. One might expect most nuclear structure effects to
change Eqs.~(\ref{pivv}, \ref{pivt}, \ref{pitt}) in similar ways.
Indeed, within a model, one could try to combine the three
polarizations and express the responses in terms of the electric
and magnetic form factors.  However, we chose to define our off-shell
currents directly from Eqs.~(\ref{jelec}, \ref{jweakv}) in
terms of $F_{\scriptscriptstyle 1}$ and $F_{\scriptscriptstyle 2}$ form
factors.
Indeed, we find (in the next section) different relativistic
effects on $\Pi^{\scriptscriptstyle \mu\nu}_{\scriptscriptstyle {\rm vv}}$
than on $\Pi^{\scriptscriptstyle \mu\nu}_{\scriptscriptstyle {\rm tt}}$.
Thus, in our model, it is important to break the response up into these
three pieces (vv, vt, and tt).

The weak-axial polarization,
$\Pi^{\scriptscriptstyle \mu;\nu5}_{\scriptscriptstyle \gamma Z}$,
can be similarly written in terms of polarizations having
either a vector or a tensor vertex, and an axial-vector vertex
a=$\gamma^{\mu}\gamma^{5}$,
  \begin{mathletters}
   \begin{eqnarray}
     i\Pi^{\scriptscriptstyle \mu;\nu5}_{\scriptscriptstyle {\rm va}} &=&
     \int {d^4k\over (2\pi)^4} {\rm Tr}
     \left[
       G(k)
       \gamma^{\scriptscriptstyle \mu}
       G(k+q)
       \gamma^{\scriptscriptstyle \nu}\gamma^{\scriptscriptstyle 5}
      \right] \;, \\
     i\Pi^{\scriptscriptstyle \mu;\nu5}_{\scriptscriptstyle {\rm ta}} &=&
     \int {d^4k\over (2\pi)^4} {\rm Tr}
     \left[
       G(k)
       {i\sigma^{\scriptscriptstyle \mu\eta} (-q)_{\scriptscriptstyle
\eta}\over 2M}
       G(k+q)
       \gamma^{\scriptscriptstyle \nu}\gamma^{\scriptscriptstyle 5}
      \right] \;.
   \end{eqnarray}
  \end{mathletters}

Note, that the Green's function defined in Eq.~\ref{greens},
describes the propagation of both nucleons and antinucleons.
Hence, the above polarization tensors contain, in addition to
the usual particle-hole excitations, nucleon-antinucleon
excitations (vacuum polarization). However, vacuum polarization
has no imaginary part in the space-like region ($q>\omega$) probed
in electron scattering. Therefore, vacuum polarization makes no
contribution to the uncorrelated responses. Vacuum polarization
will, however, make an important contribution to the correlated
response that will be presented in Sec.~\ref{secresults}.

Using the above defined polarizations together with the single-nucleon
form factors leads to the following expressions for the electromagnetic
and electromagnetic-weak polarization tensors
  \begin{mathletters}
   \begin{eqnarray}
     \Pi^{\scriptscriptstyle \mu\nu(i)}_{\scriptscriptstyle \gamma\gamma} &=&
        F_{\scriptscriptstyle 1}^{\scriptscriptstyle (i)}
        F_{\scriptscriptstyle 1}^{\scriptscriptstyle (i)}
        \Pi^{\scriptscriptstyle \mu\nu(i)}_{\scriptscriptstyle {\rm vv}} +
        \left(
         F_{\scriptscriptstyle 1}^{\scriptscriptstyle (i)}
         F_{\scriptscriptstyle 2}^{\scriptscriptstyle (i)} +
         F_{\scriptscriptstyle 2}^{\scriptscriptstyle (i)}
         F_{\scriptscriptstyle 1}^{\scriptscriptstyle (i)}
        \right)
        \Pi^{\scriptscriptstyle \mu\nu(i)}_{\scriptscriptstyle {\rm vt}} +
         F_{\scriptscriptstyle 2}^{\scriptscriptstyle (i)}
         F_{\scriptscriptstyle 2}^{\scriptscriptstyle (i)}
        \Pi^{\scriptscriptstyle \mu\nu(i)}_{\scriptscriptstyle {\rm tt}} \;,
        \label{pimunuone}   \\
     \Pi^{\scriptscriptstyle \mu\nu(i)}_{\scriptscriptstyle \gamma Z}     &=&
        F_{\scriptscriptstyle 1}^{\scriptscriptstyle (i)}
        \widetilde{F}_{\scriptscriptstyle 1}^{\scriptscriptstyle (i)}
        \Pi^{\scriptscriptstyle \mu\nu(i)}_{\scriptscriptstyle {\rm vv}} +
        \left(
         F_{\scriptscriptstyle 1}^{\scriptscriptstyle (i)}
         \widetilde{F}_{\scriptscriptstyle 2}^{\scriptscriptstyle (i)} +
         F_{\scriptscriptstyle 2}^{\scriptscriptstyle (i)}
         \widetilde{F}_{\scriptscriptstyle 1}^{\scriptscriptstyle (i)}
        \right)
        \Pi^{\scriptscriptstyle \mu\nu(i)}_{\scriptscriptstyle {\rm vt}} +
         F_{\scriptscriptstyle 2}^{\scriptscriptstyle (i)}
         \widetilde{F}_{\scriptscriptstyle 2}^{\scriptscriptstyle (i)}
        \Pi^{\scriptscriptstyle \mu\nu(i)}_{\scriptscriptstyle {\rm tt}} \;,
        \label{pimunutwo}   \\
     \Pi^{\scriptscriptstyle \mu;\nu5(i)}_{\scriptscriptstyle \gamma Z}     &=&
         F_{\scriptscriptstyle 1}^{\scriptscriptstyle (i)}
         \widetilde{G}_{\scriptscriptstyle A}^{\scriptscriptstyle (i)}
        \Pi^{\scriptscriptstyle \mu;\nu5(i)}_{\scriptscriptstyle {\rm va}} +
         F_{\scriptscriptstyle 2}^{\scriptscriptstyle (i)}
         \widetilde{G}_{\scriptscriptstyle A}^{\scriptscriptstyle (i)}
        \Pi^{\scriptscriptstyle \mu;\nu5(i)}_{\scriptscriptstyle {\rm ta}} \;.
        \label{pimunuthree}
    \end{eqnarray}
   \end{mathletters}

To summarize, the response functions depend on a variety of nucleon
electromagnetic and weak form factors. These form factors are collected
in the Appendix.  The nuclear structure, on the other hand, is contained
in the polarization insertions
$\Pi^{\scriptscriptstyle \mu\nu}_{\scriptscriptstyle {\rm vv}}\;,
 \Pi^{\scriptscriptstyle \mu\nu}_{\scriptscriptstyle {\rm vt}}\;,
 \Pi^{\scriptscriptstyle \mu\nu}_{\scriptscriptstyle {\rm tt}}\;,
 \Pi^{\scriptscriptstyle \mu;\nu5}_{\scriptscriptstyle {\rm va}}\;,$ and
$\Pi^{\scriptscriptstyle \mu;\nu5}_{\scriptscriptstyle {\rm ta}}\;$.

The response functions can be calculated in various approximations.
The simplest approximation is to calculate uncorrelated response
functions for nuclear matter. This can be done either for $M^*=M$,
which gives the free Fermi gas results, or for a smaller value of $M^*$.
These results were presented in Ref.~\cite{horow92} and are included
below for comparison. Note that in these approximations the isoscalar
and isovector polarizations are identical. The change in the responses
arises exclusively from the difference in isoscalar and isovector
single-nucleon form factors.

Alternatively, one can calculate the response functions directly in a
finite nucleus and avoid a local density approximation. The finite-nucleus
calculations can be carried out using a nonspectral representation of the
nucleon Green's function as discussed in
Refs.~\cite{shep88,wehbec88,horpie90}. Next, long range correlations can
be included by calculating RPA polarization insertions
Refs.~\cite{shep88,wehbec88,horpie90,shep89}. We employ a residual
interaction consisting of only isoscalar sigma and omega meson exchange
with self-consistent parameters (see Table~\ref{tablemft}) chosen to fit
ground-state properties~\cite{horser81}.
In particular note that the isovector
response is insensitive to correlations and thus remains identical
to its uncorrelated value. Finally, vacuum polarization
can be included, in addition to particle-hole or core polarization,
in calculating the RPA response. Further details can be found
in Ref.~\cite{horpie90}.

\section{Results}
\label{secresults}

We now present response functions for the N=Z target $^{12}$C. We
expect similar results for heavier nuclei. For simplicity we consider
only a momentum transfer of $q=550\;$MeV/c. Traditional nuclear structure
uncertainties will be larger for much smaller momentum transfers.
We expect similar relativistic effects for momentum transfers greater
than $q=550\;$MeV/c, but this remains to be investigated.

First we notice that the parity-violating asymmetry ${\cal A}$ is dominated
by the transverse weak vector contribution ${\cal A}_{\scriptscriptstyle T}$
with
${\cal A}_{\scriptscriptstyle T'}$ making perhaps a 20 percent contribution for
large scattering angles and ${\cal A}_{\scriptscriptstyle L}$ contributing
perhaps
10 percent at forward angles. Furthermore, ${\cal A}_{\scriptscriptstyle T}$
is itself dominated by the isovector response from the
${F}_{\scriptscriptstyle 2}^{\scriptscriptstyle (1)}$ form factor contributions
(last term in Eq.~(\ref{pimunutwo})).

For a large-angle experiment, say at an electron scattering angle
of 150 degrees, ${\cal A}_{\scriptscriptstyle L}$ is unimportant (as is the
contribution from the longitudinal response $S_{\scriptscriptstyle L}$ to the
parity-conserving cross section) because of the kinematic factors
$v_{\scriptscriptstyle L}$ and $v_{\scriptscriptstyle T}$
($v_{\scriptscriptstyle L}/v_{\scriptscriptstyle T}
\rightarrow 0$ at backward angles). Thus, let us focus on
${\cal A}_{\scriptscriptstyle T}$. We will return to the small
${\cal A}_{\scriptscriptstyle T'}$ contribution at the end. Furthermore,
we will neglect strange-quark contributions
so that, from Eqs.~(\ref{ftildea} and  \ref{ftildeb}),
   $\widetilde{F}_{\scriptscriptstyle i}^{\scriptscriptstyle (0)}=
    \sqrt{3}\;\xi_{\scriptscriptstyle V}^{\scriptscriptstyle (0)}
              {F}_{\scriptscriptstyle i}^{\scriptscriptstyle (0)}$
and
   $\widetilde{F}_{\scriptscriptstyle i}^{\scriptscriptstyle (1)}=
              \xi_{\scriptscriptstyle V}^{\scriptscriptstyle (1)}
              {F}_{\scriptscriptstyle i}^{\scriptscriptstyle (1)}$.
Thus ${\cal A}_{\scriptscriptstyle T}$ only depends on the ratio of
isoscalar $S_{\scriptscriptstyle T}^{\scriptscriptstyle (0)}$ to isovector
$S_{\scriptscriptstyle T}^{\scriptscriptstyle (1)}$ electromagnetic responses.
The only nuclear structure that ${\cal A}_{\scriptscriptstyle T}$ is
sensitive to is isospin. At large angles then, ${\cal A}_{\scriptscriptstyle
T}$
becomes (neglecting the contribution from strange quarks)
  \begin{equation}
   {\cal A}_{\scriptscriptstyle T}=
   g_{\scriptscriptstyle A}^{\scriptscriptstyle e}
   \xi_{\scriptscriptstyle V}^{\scriptscriptstyle (1)}
   \left[
    {1 + {\sqrt{3}\;\xi_{\scriptscriptstyle V}^{\scriptscriptstyle (0)} \over
                                \xi_{\scriptscriptstyle V}^{\scriptscriptstyle
(1)}}
                                {\cal R}_{\scriptscriptstyle T} \over
     1 + {\cal R}_{\scriptscriptstyle T}}
    \right] \;,
   \label{transa}
  \end{equation}
where the isospin ratio is given by,
  \begin{equation}
   {\cal R}_{\scriptscriptstyle T} \equiv
    S_{\scriptscriptstyle T}^{\scriptscriptstyle (0)}/
    S_{\scriptscriptstyle T}^{\scriptscriptstyle (1)} \;.
   \label{tratio}
  \end{equation}
The couplings $\sqrt{3}\;\xi_{\scriptscriptstyle V}^{\scriptscriptstyle (0)}$
and
$\xi_{\scriptscriptstyle V}^{\scriptscriptstyle (1)}$ (Table~\ref{tablexis})
have
opposite signs and almost equal magnitude. Therefore a change
in the ratio ${\cal R}_{\scriptscriptstyle T}$ by 0.01 ({\it i.e.,}
${\cal R}_{\scriptscriptstyle T} \rightarrow {\cal
R}_{\scriptscriptstyle T}\pm 0.01$)
will change
${\cal A}_{\scriptscriptstyle T}$ by about two percent. However, the total
transverse response
($S_{\scriptscriptstyle T}^{\scriptscriptstyle
(0)}+S_{\scriptscriptstyle T}^{\scriptscriptstyle (1)}$)
is predominantly isovector because of the large isovector anomalous
moment of the nucleon. Thus ${\cal R}_{\scriptscriptstyle T}$ is expected to be
small (of the order of 5\%). However, one must investigate the nuclear
structure uncertainties in ${\cal R}_{\scriptscriptstyle T}$. Most traditional
corrections to the isoscalar response are hidden because of the small
isoscalar anomalous moment of the nucleon.

However, there could be large relativistic corrections to the isoscalar
transverse response $S_{\scriptscriptstyle T}^{\scriptscriptstyle
(0)}$. A smaller value of
$M^*$\ enhances $\Pi_{\scriptscriptstyle {\rm vv}}^{11}$ and thus increases
$S_{\scriptscriptstyle T}^{\scriptscriptstyle (0)}$. [Note,
$S_{\scriptscriptstyle T}^{\scriptscriptstyle (0)}$ is
dominated by $\Pi_{\scriptscriptstyle {\rm vv}}^{\scriptscriptstyle
11}$ rather than by
$\Pi_{\scriptscriptstyle {\rm tt}}^{\scriptscriptstyle 11}$ because of
the small value for
${F}_{\scriptscriptstyle 2}^{\scriptscriptstyle (0)}$.] The smaller
value of $M^*$ enhances
a nucleon's velocity and produces a larger convection current
(which is half isoscalar). This could lead to a large change in
${\cal A}_{\scriptscriptstyle T}$.

Below, we present results showing that the uncertainties in the
transverse response, and hence on ${\cal A}_{\scriptscriptstyle T}$, are large.
For simplicity we do not show results for longitudinal responses. However,
most of these have already been published in Ref.~\cite{horpie90}. It is
already generally agreed that nuclear structure uncertainties in the
longitudinal response (and hence in ${\cal A}_{\scriptscriptstyle L}$) are
large.

We present a catalog of transverse responses for ten different
calculations. First, we consider four different finite nucleus
calculations. Uncorrelated (or impulse) and RPA responses are
shown first neglecting vacuum polarization in a so called mean
field theory (MFT) and then with vacuum polarization in a
relativistic Hartree approximation (RHA). For comparison we
also show the corresponding four responses calculated in nuclear
matter.  This allows one to examine commonly made local density
approximations.  Finally, we also show two ``nonrelativistic''
calculations; one with and the other without RPA correlations.
Here, a Fermi gas is considered with an effective mass equal to
the free nucleon mass. In Table~\ref{tablemft} we list both sets
of parameters used in the calculations.

Ten calculations lead to somewhat complicated results.  However, there
are important motivations to examine RPA, vacuum polarization and full
finite nucleus corrections to simple local density approximation results.
First, relativistic effects for magnetic moments, seen in impulse
approximations, vanish in full RPA calculations~\cite{mcneil86,furser87}.
Do the relativistic effects for ${\cal A}_{\scriptscriptstyle T}$
claimed in Ref.~\cite{horow92} also vanish in RPA?  Below, we see that
they do not.

Next, vacuum polarization may have an important effect on the unpolarized
longitudinal response~\cite{horpie90,horow88,kursuz88}.
As the effective nucleon mass decreases it
is easier to excite virtual nucleon-antinucleon pairs out of the
vacuum.  Therefore, the vacuum becomes better at screening
charge.  This change in vacuum polarization with $M^*$\ reduces
the charge visible to the electron probe.  Relativistic RPA
calculations including vacuum polarization have longitudinal responses
reduced by about twenty five percent compared to impulse approximations. This
reduction is in good agreement with experiment for medium and heavy nuclei
(although the experimental errors are large).  In contrast,
nonrelativistic calculations have integrated responses constrained
by the Coulomb sum rule and over predict the data. Below we examine
the effect of vacuum polarization on parity violation.

Finally, many quasielastic calculations are done in a local density
approximation (LDA). This is expected to be qualitatively correct for the
high excitation energies of quasifree scattering.  However, we check
to see if parity violation is sensitive to any of the details of the
response not reproduced by a LDA. In particular the high and low
excitation energy parts of the response are poorly described in LDA.

Figure~{\ref{figone}} shows transverse response functions calculated
in nuclear matter compared to data from $^{12}$C at a momentum transfer
of $q=550$~MeV/c. We consider an average density of $^{12}$C which
corresponds to a Fermi momentum of $k_{\scriptscriptstyle F}=225$~MeV/c. First,
free Fermi gas results are shown with $M^*=M$. These are equivalent
to the results reported in Ref.~\cite{donnel92}. Next, relativistic
MFT results are shown, as in Ref.~\cite{horow92}, with an
effective mass of $M^*=0.68M$. This is the self-consistent value for
the MFT at a density corresponding to $k_{\scriptscriptstyle F}=225$~MeV/c. The
primary effect of $M^*$ is to shift the strength to higher energies in
agreement with data. Of course there is still substantial experimental
strength at medium to high excitation energies. Presumably this
corresponds to delta and pion production, multinucleon knockout and
or meson exchange currents. These effects are outside the scope of the
one-body mechanism assumed for all of our calculations. We should
mention that no attempt was made to fit the data by selecting an
optimal value for $k_{\scriptscriptstyle F}$. As we will show below
finite-nucleus
calculations give a more realistic distribution of transverse strength.

The contribution of other mechanisms to the quasifree response is an
important source of uncertainty for parity violating experiments.
At this time it is not well known the relative importance of different
processes much less their isoscalar vs. isovector contribution.
One might argue that Delta or pion exchange currents are isovector.
However, there could be isoscalar contributions from other mesons or
from two-nucleon correlations.  Any effect from these
other contributions will only strengthen our conclusion that
there are important nuclear structure uncertainties in parity
violating quasielastic electron scattering.

We also show RPA responses in Fig.~{\ref{figone}}. These use a simple
isoscalar interaction consisting of sigma and omega meson exchange. The
parameters are taken from the MFT fit to the saturation density and
binding energy of nuclear matter~\cite{waleck74,serwal86}.
We assume that the isovector interaction is zero. Consequently,
since the total response is dominated by the isovector part, RPA
has almost no effect on the unpolarized transverse response.

Next, in Fig.~{\ref{figtwo}} we show the isoscalar contribution to
the transverse response. In principle, this response contributes to
the observable unpolarized cross section. However, because it is so
small (compare the scales of Figs.~{\ref{figone}} and {\ref{figtwo}})
it is essentially unconstrained by data. As $M^*$ decreases the
isoscalar response is predicted to be significantly enhanced.
This effect was discussed in Ref.~\cite{horow92}.

The RPA correlations are seen to shift strength downward (due to an
attractive particle-hole interaction) and somewhat reduce the area.
This is true in both the MFT and $M^*=M$ calculations. Note, there
is no simple relation between the MFT RPA and the free Fermi-gas results.
This is in contrast to the situation for (isoscalar) magnetic
moments where there was a Fermi liquid theory theorem constraining the
results~\cite{mcneil86,furser87}. Instead, the quasifree response at
finite momentum transfer is a dynamical quantity and depends on the details
of the model. There is no theorem which constrains the finite-$q$ response.

The ratio ${\cal R}_{\scriptscriptstyle T}$ of isoscalar to isovector
transverse
responses is shown in Fig.~{\ref{figthree}}. This is the only nuclear
structure information needed to determine ${\cal A}_{\scriptscriptstyle T}$
(see Eq.~(\ref{transa})). In general there is a large enhancement in
this ratio for the MFT calculations. However, there are also large
effects from RPA correlations in both the MFT and $M^*=M$ calculations.
Figure~{\ref{figthree}} also shows ${\cal A}_{\scriptscriptstyle T}$ deduced
from
${\cal R}_{\scriptscriptstyle T}$ by using Eq.~(\ref{transa}). This quantity is
directly related to the experimental asymmetry. The 5 to 10 percent
spread in ${\cal A}_{\scriptscriptstyle T}$ is much larger than the perhaps one
to two percent accuracy~\cite{donnel92} needed to measure strange
quark effects or radiative corrections to the axial contributions.

We now consider full finite nucleus calculations for $^{12}$C.
Our finite nucleus calculations have been described in
Ref.~\cite{horpie90} and are based on calculating the nucleon
Green's function directly in the finite system. The RPA integral
equations are then solved by matrix inversion in momentum space. The
interaction parameters are the same ones used to fit ground-state
properties so the calculations explicitly conserve current.

Figure~{\ref{figfour}} shows MFT results for the total transverse response.
First, the finite nucleus calculations are qualitatively similar to the
nuclear matter results. However, the shift in the position of the peak
and the broadening of the width are not as pronounced as in the nuclear
matter case. In addition, there is a high energy tail in the finite
nucleus response which is not reproduced by the nuclear matter
calculations. Figure~{\ref{figfive}} shows isoscalar responses.
Again, the finite nucleus responses have a high energy tail not
present in nuclear matter. However, the LDA results are qualitatively
similar for the lower energies.

The isospin ratio ${\cal R}_{\scriptscriptstyle T}$ is shown in
Fig.~{\ref{figsix}}.
This is qualitatively similar in finite nucleus and LDA calculations.
However, the uncorrelated finite nucleus calculations predict an
enhancement in ${\cal R}_{\scriptscriptstyle T}$ at high excitation energies.
This is due to the different energy dependences of
$\Pi_{\scriptscriptstyle {\rm vv}}^{\scriptscriptstyle 11}$ (which is
dominant for the
isoscalar response) and
$\Pi_{\scriptscriptstyle {\rm tt}}^{\scriptscriptstyle 11}$ (which
dominates the isovector)
in the finite
nucleus results. Note, we have not yet investigated the origin of this
difference. The ${\cal A}_{\scriptscriptstyle T}$ deduced from these ratios are
also shown in Fig.~{\ref{figsix}}. Again there is a 5 to 10
percent dispersion in the results.

Finally, we consider RHA calculations which include vacuum
polarization~\cite{horpie90,horow88}. Both nuclear matter and
finite nucleus RHA results are shown in
Figs.~({\ref{figseven}, {\ref{figeight}, {\ref{fignine}).
The most important difference between these curves and MFT
results may not be intrinsic to vacuum polarization.  Instead, the RHA has a
somewhat larger effective mass.  This is because the RHA includes a
term describing the change in the energy density of the vacuum (with $M^*$).
This term tends to resist changes in the nucleon mass. [Also, the RHA
reproduces nuclear matter saturation with somewhat smaller meson couplings].
Therefore, RHA results tend to be intermediate between MFT and
$M^*=M$\ calculations. In particular, the isoscalar response is
not enhanced as much in the RHA as it is in the MFT.

In summary we have presented results for ${\cal A}_{\scriptscriptstyle T}$
for ten different calculations in
Figs.~({\ref{figthree}, {\ref{figsix} and {\ref{fignine}).
Depending on the approximation uncorrelated vs. RPA or MFT vs. RHA
one obtains substantially different results. The five to ten percent
dispersion in ${\cal A}_{\scriptscriptstyle T}$ is larger than the accuracy
needed to extract strange quark information.

For completeness we also show axial response functions
and the ${\cal A}_{\scriptscriptstyle T'}$ derived from them in
Fig.~{\ref{figten}}. Since we have ignored strange-quark
contributions, the axial response is purely isovector. Thus,
(isoscalar) RPA correlations have no effect on the response
and we only report uncorrelated results.

We see that the position and width of the quasielastic peak
vary considerably reflecting the different effective masses
of the various models. In addition, the finite nucleus calculations
show their characteristic high-energy tails. Again, we observe
differences of the order of 5 to 10 percent in the results.

Finally, in
Figs.~({\ref{figeleven}}, {\ref{figtwelve}}, and {\ref{figthirteen}})
we simply plot all ten predictions for the total asymmetry
${\cal A}={\cal A}_{\scriptscriptstyle L}+
          {\cal A}_{\scriptscriptstyle T}+
          {\cal A}_{\scriptscriptstyle T'}$.
These show the total dispersion in the different results.
The experimental asymmetry is simply
$a_{\scriptscriptstyle 0}\tau$ times these curves. At $q=550$~MeV/c and at
an excitation energy of $\omega=150$~MeV the overall magnitude
is $a_{\scriptscriptstyle 0}\tau=-2.52\times10^{-5}$
(see also Table~\ref{tableasym}).

\section{Summary and Conclusions}
\label{secconcl}

In principle, a parity violation measurement in quasielastic electron
scattering can provide information on strange quark
contributions and radiative corrections to weak axial currents.
This could greatly enhance the value of an elastic electron-proton
experiment. However, in order for this to be useful one must understand
nuclear structure corrections to the quasielastic measurement.

The quasielastic measurement is primarily sensitive to the ratio of
isoscalar to isovector transverse responses ${\cal R}_{\scriptscriptstyle T}$.
We have calculated these responses in a variety of relativistic mean
field and RPA models for $^{12}$C and nuclear matter at a momentum
transfer of $q=550$~MeV/c.  We find that ${\cal R}_{\scriptscriptstyle T}$
is sensitive to relativistic effects from a strong scalar
field (small $M^*$) and from RPA correlations. Furthermore, full finite
nucleus calculations predict large changes in ${\cal R}_{\scriptscriptstyle T}$
in the high energy tail of the response.

These nuclear structure effects could change the asymmetry
${\cal A}$ by as much as ten percent. This uncertainty is
so large that a quasielastic measurement may not provide useful
information on strange quark contributions or radiative corrections
(unless these effects can be significantly constrained by future work).
We note that a quasielastic measurement from deuterium rather than
from a heavier nucleus should not be plagued by these uncertainties.
We expect relativistic corrections to be much smaller for deuterium.

Alternatively, a quasielastic parity violating measurement could
provide information on relativistic and nuclear structure effects
on the isoscalar transverse response. This small response is interesting
and not well determined by other data. This information might be obtained
by comparing a deuterium measurement to that from a heavier nucleus such
as $^{12}$C.

\acknowledgments
This research was supported by the Florida
State University Supercomputer Computations Research Institute and
U.S. Department of Energy contracts DE-FC05-85ER250000,
DE-FG05-92ER40750, and DE-FG02-87ER40365.

\unletteredappendix{Nucleon Form Factor}
\label{appx}

We adopt the form factor parameterization used in Ref.~\cite{donnel92}.
First the electromagnetic form factors are written in terms of a simple
dipole,
  \begin{eqnarray}
    G &=& (1+4.97\tau)^{-2}                       \;, \\
    F_{\scriptscriptstyle 1}^{\scriptscriptstyle (p)} &=&
    \left[ 1+\tau(1+\lambda_p) \right]G/(1+\tau)  \;, \\
    F_{\scriptscriptstyle 2}^{\scriptscriptstyle (p)} &=&
    \lambda_p G/(1+\tau)                          \;, \\
    F_{\scriptscriptstyle 1}^{\scriptscriptstyle (n)} &=&
    \tau\lambda_n(1-\eta)G/(1+\tau)               \;, \\
    F_{\scriptscriptstyle 2}^{\scriptscriptstyle (n)} &=&
    \lambda_n(1+\tau\eta)G/(1+\tau)               \;.
  \end{eqnarray}
Here the anomalous moments are,
 \begin{equation}
   \lambda_p=1.793 \;, \qquad \lambda_n=-1.913 \;,
 \end{equation}
and
 \begin{equation}
  \eta=(1+5.6\tau)^{-1} \;.
 \end{equation}
This parameterization is good for the neutron form factors
provided $\tau \ll 1$. Finally, the isovector axial form factor
is ($g_a=$1.26)
 \begin{equation}
   G_{\scriptscriptstyle A}^{\scriptscriptstyle (3)} =
     {g_a \over 2} (1+3.53\tau)^{-2} \;.
 \end{equation}

\figure{Transverse response as a function of energy loss
        for ${}^{12}$C at $q=550$~MeV. The solid (dot-dashed) line
        gives the uncorrelated (RPA) Fermi gas result. The dashed
        (dotted) line gives the uncorrelated (RPA) result using
        an effective mass of $M^*=0.68$. All calculations
        ignore vacuum polarization and used a Fermi momentum of
        $k_{\scriptscriptstyle F}=225$~MeV. Experimental data is from
        Ref.~\cite{mezian84}.
        \label{figone}}
\figure{Same as Fig.~\ref{figone} but for the isoscalar
        contribution to the transverse response.
        \label{figtwo}}
\figure{Same as Fig.~\ref{figone} but for: (a) the transverse
        isoscalar to isovector ratio and (b) the transverse
        contribution to the parity-violating asymmetry at
        $\theta=150^{\circ}$.
        \label{figthree}}
\figure{Transverse response as a function of energy loss
        for ${}^{12}$C at $q=550$~MeV. The solid (dot-dashed) line
        gives the uncorrelated (RPA) finite-nucleus result. The
        dashed (dotted) line gives the uncorrelated (RPA) result
        using an effective mass of $M^*=0.68$. All calculations
        ignore vacuum polarization and (the nuclear matter ones)
        used a Fermi momentum of $k_{\scriptscriptstyle F}=225$~MeV.
        Experimental data is from Ref.~\cite{mezian84}.
        \label{figfour}}
\figure{Same as Fig.~\ref{figfour} but for the isoscalar
        contribution to the transverse response.
        \label{figfive}}
\figure{Same as Fig.~\ref{figfour} but for: (a) the transverse
        isoscalar to isovector ratio and (b) the transverse
        contribution to the parity-violating asymmetry at
        $\theta=150^{\circ}$.
        \label{figsix}}
\figure{Transverse response as a function of energy loss
        for ${}^{12}$C at $q=550$~MeV. The solid (dot-dashed) line
        gives the uncorrelated (RPA) finite-nucleus result. The
        dashed (dotted) line gives the uncorrelated (RPA) result
        using an effective mass of $M^*=0.80$. All calculations
        include vacuum polarization and (the nuclear matter ones)
        used a Fermi momentum of $k_{\scriptscriptstyle F}=225$~MeV.
        Experimental data is from Ref.~\cite{mezian84}.
        \label{figseven}}
\figure{Same as Fig.~\ref{figseven} but for the isoscalar
        contribution to the transverse response.
        \label{figeight}}
\figure{Same as Fig.~\ref{figseven} but for: (a) the transverse
        isoscalar to isovector ratio and (b) the transverse
        contribution to the parity-violating asymmetry at
        $\theta=150^{\circ}$.
        \label{fignine}}
\figure{Axial isovector response (a), and its contribution to the
        parity-violating asymmetry at $\theta= 150^{\circ}$
        (b), as a function of energy loss for ${}^{12}$C at
        $q=550$~MeV. The solid (dot-dashed) line gives the
        MFT (RHA) finite-nucleus result. The dashed (dotted) line
        gives the MFT (RHA) nuclear matter result. Also shown
        are free Fermi gas results. Nuclear matter calculations
        used a Fermi momentum of $k_{\scriptscriptstyle F}=225$~MeV.
        \label{figten}}
\figure{Parity-violating asymmetry as a function of energy loss
        for ${}^{12}$C at $\theta=150^{\circ}$ and
        $q=550$~MeV. The solid (dot-dashed) line
        gives the uncorrelated (RPA) Fermi gas result. The dashed
        (dotted) line gives the uncorrelated (RPA) result using
        an effective mass of $M^*=0.68$. All calculations
        ignore vacuum polarization and used a Fermi momentum of
        $k_{\scriptscriptstyle F}=225$~MeV.
        \label{figeleven}}
\figure{Parity-violating asymmetry as a function of energy loss
        for ${}^{12}$C at $\theta=150^{\circ}$ and
        $q=550$~MeV. The solid (dot-dashed) line
        gives the uncorrelated (RPA) finite-nucleus result. The
        dashed (dotted) line gives the uncorrelated (RPA) result
        using an effective mass of $M^*=0.68$. All calculations
        ignore vacuum polarization and (the nuclear matter ones) used
        a Fermi momentum of $k_{\scriptscriptstyle F}=225$~MeV.
        \label{figtwelve}}
\figure{Parity-violating asymmetry as a function of energy loss
        for ${}^{12}$C at $\theta=150^{\circ}$ and
        $q=550$~MeV. The solid (dot-dashed) line
        gives the uncorrelated (RPA) finite-nucleus result. The
        dashed (dotted) line gives the uncorrelated (RPA) result
        using an effective mass of $M^*=0.80$. All calculations
        include vacuum polarization and (the nuclear matter ones) used
        a Fermi momentum of $k_{\scriptscriptstyle F}=225$~MeV.
        \label{figthirteen}}

 \mediumtext
 \begin{table}
  \caption{Electro-weak charges for electrons and $u$, $d$, and, $s$
           quarks. Numerical values were evaluated using
           $\sin^{2}\theta_{\scriptscriptstyle W}\simeq0.227$.}
   \begin{tabular}{cccc}
    Fermion & $e(EM)$ & $g_{\scriptscriptstyle V}$(Weak Vector)  &
                        $g_{\scriptscriptstyle A}$(Weak Axial)   \\
        \tableline
    $e$ & $-1  $ &
    $-1+4\sin^{2}\theta_{\scriptscriptstyle W}
    \simeq -0.091$ & $+1$                                      \\
    $u$ & $+2/3$ &
    $+1-{{8 \over 3}}\sin^{2}\theta_{\scriptscriptstyle W}
    \simeq +0.394$ & $-1$                                       \\
    $d$ & $-1/3$ &
    $-1+{{4 \over 3}}\sin^{2}\theta_{\scriptscriptstyle W}
    \simeq -0.697$ & $+1$                                       \\
    $s$ & $-1/3$ &
    $-1+{{4 \over 3}}\sin^{2}\theta_{\scriptscriptstyle W}
    \simeq -0.697$ & $+1$                                       \\
   \end{tabular}
  \label{tablegs}
 \end{table}

 \mediumtext
 \begin{table}
  \caption{Isoscalar, isovector, and strange (or singlet) weak-vector
           and weak-axial couplings. Numerical values were evaluated
           using $\sin^{2}\theta_{\scriptscriptstyle W}\simeq0.227$.}
   \begin{tabular}{ccc}
    Coupling   &   (Weak Vector)   &   (Weak Axial)       \\
        \tableline
    $\sqrt{3}\;\xi^{\scriptscriptstyle (0)}$  &
    $-4\sin^{2}\theta_{\scriptscriptstyle W}\simeq -0.909$ & $0$     \\
    $\xi^{\scriptscriptstyle (1)}$            &
    $2-4\sin^{2}\theta_{\scriptscriptstyle W}\simeq +1.091$ & $-2$   \\
    $\xi^{\scriptscriptstyle (s)}$            &
    $-1$ & $+1$                                           \\
   \end{tabular}
  \label{tablexis}
 \end{table}

 \mediumtext
 \begin{table}
  \caption{Coupling constants, meson masses, and effective
           nucleon mass in a mean-field approximation to
           the Walecka model. The effective nucleon mass
           was calculated using a Fermi momentum of
           $k_{\scriptscriptstyle {\rm F}}=225$~MeV.}
   \begin{tabular}{cccccc}
    Model &  $g_{\scriptscriptstyle {\rm s}}^{\scriptscriptstyle 2}$  &
             $g_{\scriptscriptstyle {\rm v}}^{\scriptscriptstyle 2}$  &
             $m_{\scriptscriptstyle {\rm s}}$~(MeV)        &
             $m_{\scriptscriptstyle {\rm v}}$~(MeV)        &
             $M^{*}/M$                          \\
        \tableline
    MFT & 109.626  & 190.431  & 520 & 783 & 0.68  \\
    RHA &  54.289  & 102.770  & 458 & 783 & 0.80  \\
   \end{tabular}
  \label{tablemft}
 \end{table}

 \mediumtext
 \begin{table}
  \caption{Ratio of isoscalar to isovector transverse responses and
           parity-violating asymmetries for ${}^{12}$C using various
           nuclear-structure models at $q=550$~MeV, $\omega=150$~MeV,
           and $\theta=150^\circ$. The nuclear matter calculations
           used a Fermi momentum of $k_{\scriptscriptstyle {\rm F}}=225$~MeV.
           Within each model, the first (second) row of numbers give
           the uncorrelated (RPA) results.}
   \begin{tabular}{ccccc}
          Model      &  ${\cal R}_{\scriptscriptstyle T}$  &
          ${\cal A}_{\scriptscriptstyle T}/a_{\scriptscriptstyle 0}\tau$     &
          ${\cal A}_{\scriptscriptstyle T'}/a_{\scriptscriptstyle 0}\tau$    &
          ${\cal A}/a_{\scriptscriptstyle 0}\tau$                \\
        \tableline
    Free Fermi Gas        &       0.044       &       0.973
                          &       0.198       &       1.168       \\
                          &       0.030       &       1.000
                          &       0.201       &       1.201       \\
        \tableline
    MFT$[M^{\ast}=0.68]$  &       0.083       &       0.915
                          &       0.180       &       1.090       \\
                          &       0.071       &       0.937
                          &       0.182       &       1.117       \\
        \tableline
    MFT [Fin. Nuc.]       &       0.072       &       0.928
                          &       0.184       &       1.107       \\
                          &       0.068       &       0.937
                          &       0.185       &       1.119       \\
        \tableline
    RHA$[M^{\ast}=0.80]$  &       0.064       &       0.943
                          &       0.188       &       1.126       \\
                          &       0.039       &       0.994
                          &       0.194       &       1.191       \\
        \tableline
    RHA [Fin. Nuc.]       &       0.054       &       0.957
                          &       0.192       &       1.144       \\
                          &       0.051       &       0.969
                          &       0.194       &       1.164       \\
   \end{tabular}
  \label{tableasym}
 \end{table}

\end{document}